\documentstyle[aaspp,psfig]{article}
\newcommand{\etal}{et al.}

\newcommand{\mn}{MNRAS}

\newcommand{\aandasupp}{A\&AS}
\newcommand{\aanda}{A\&A}

\newcommand{\half}{\mbox{$\frac{1}{2}$}}

\newcommand{\gtn}{\mbox{G29.96$-$0.02}}
\newcommand{\degs}{\mbox{$^{\mbox{o}}$}}

\newcommand{\kms}{\mbox{km s$^{-1}$}}
\newcommand{\Wm}{\mbox{W m$^{-2}$}}

\newcommand{\brg}{\mbox{Br$\gamma$}}

\begin{document}

\title{An IR study of the velocity structure of the cometary compact H
II region G29.96-0.02}
\author{Stuart L. Lumsden}
\affil{Anglo-Australian Observatory, PO Box 296, Epping NSW 2121, Australia}

\and

\author{Melvin G. Hoare}
\affil{Max-Planck-Institut f\"{u}r Astronomie, K\"{o}nigstuhl 17,
D-69117, Heidelberg, Germany.}

\begin{abstract}

We have mapped the velocity structure of the cometary compact H~II
region \gtn\ using long-slit echelle spectra of the \brg\ line. This
technique detects line emission over a much wider area at the
necessary spatial resolution compared to radio recombination line
observations.  Significant structure in both the velocity centroids
and the line widths is seen over the entire nebula.  Large line widths
are seen ahead of the bow and in the tail which may be due to
turbulent motions in shocked and interface regions respectively.  We
construct analytic models of the density and velocity structure in
order to attempt to distinguish between the bow shock and champagne
flow models which have been put forward to explain the cometary
morphology of many compact H~II regions. The bow shock model is unable
to explain the large velocity gradient that we see right across the
tail of the cometary region which can only be explained by the
streaming motions towards low density regions in the champagne
model. However, our approximation to the champagne model is also not
able to fit all of the features of the data. More realistic versions
of this model which include the effects of stellar winds and density
gradients may be able to provide a better match to these data.

\end{abstract}


\section{INTRODUCTION}
Young hot stars have a dramatic effect on their surroundings when they
first begin to ionize the ambient molecular gas to form H~II regions.
These effects are perhaps most important in their relation to further
star formation; do they trigger further collapse through compression
(Elmegreen \& Lada 1977) or ultimately halt the process due to
dispersal of the molecular cloud (e.g. Franco \etal\ 1994)?  The most
compact and heavily embedded sources have attracted particular
attention since they have been presumed to be the youngest and
therefore most relevant to high mass star formation. These small,
dense regions are easily visible across great distances at radio
wavelengths via their free-free emission and at far infrared
wavelengths since the hot stars also heat the associated dust.

The study of ultracompact (UC) H~II regions in particular has been stimulated
in recent years by the sparsely sampled VLA survey of Wood \& Churchwell
(1989).   One of the greatest puzzles of the Wood \& Churchwell survey is the
total number of such regions found.   By taking both the total number of O
stars in the galaxy (simply from extrapolating the number of optically visible
OB associations), and their lifetimes on the main sequence, they predicted the
total number of such sources that might still be expected to be in the UC H~II
phase.  Comparing this with the numbers actually found  led Wood \& Churchwell
to conclude that there was an order of magnitude more sources in their VLA
survey than there should have been. Since there is no other evidence for the
massive star formation rate being this much larger than expected,  the
likeliest solution to this problem is  that some mechanism is constraining the
outward expansion of the H~II region.  Wood \& Churchwell estimate that the UC
H~II  phase must last about ten times longer than the $\sim10^4$
years predicted on the
basis of simple Str\"{o}mgren sphere expansion.

Another key result from the Wood \& Churchwell study is that about 20\% of
these regions have a `cometary' appearance.  A subsequent survey by
Kurtz \etal\ (1994) and a VLA survey of much larger, optically visible,
H~II regions by Fich (1993) also found a similar proportion of this
type. The near parabolic shape of many of these sources lead Wood \&
Churchwell to develop a bow-shock interpretation for UC H~II regions
which provides an explanation for both of the lifetime of the UC
H~II region phase and the cometary morphologies.
The basis of this model is simple.  An OB star moves supersonically
through a molecular cloud ($v_*>0.2$~\kms) and the stellar wind
from the star supports a bow shock along its direction of motion.
This shock can trap the ionisation front (IF), preventing it from
expanding.  Such a situation is potentially stable, and the lifetime
of the UC H~II phase in this model is simply the star's crossing time
through the cloud, typically of the order of 10$^5$ years.

Cometary-like morphologies for H~II regions have been known for some
time and were originally labelled `blisters' by Israel (1978) due to
their propensity for being found near the edges of molecular
clouds. Icke, Gatley \& Israel (1980) developed the idea that if there
is a density gradient in the ambient gas the H~II region will expand
fastest in the low density direction and so become very
asymmetric. Tenorio-Tagle and co-workers in a series of papers
(Tenorio-Tagle 1979; Bodenheimer, Tenorio-Tagle \& Yorke 1979;
Tenorio-Tagle, Yorke \& Bodenheimer 1979; Yorke, Tenorio-Tagle \&
Bodenheimer 1983, Yorke, Tenorio-Tagle \& Bodenheimer 1984) examined
the gas dynamics of this situation and found that the pressure
gradient set up when the IF reaches the edge of the cloud causes a
`champagne' flow of ionized gas away from the cloud with velocities of
order 30 \kms.  They also give predictions for line profiles, and for
the expected radio continuum. This model does not answer the lifetime
problem however since there is no constraint on the H~II region
expansion.  Crucially, it also does not include the effects of the
stellar wind from the OB star.  Turner \& Matthews (1984) considered
the latter problem in a uniform static configuration: they find the IF
could be trapped at early times in the shell formed by the stellar
wind sweeping up the dense ambient gas ($n_e\sim10^5$~cm$^{-3}$). A
combination of this effect and the blister geometry plausibly
satisfies both the lifetime and morphology constraints.  The combined
model would have a shell structure near the core, and diffuse emission
near the tail where lower densities preclude the trapping of the IF.

The bow-shock and champagne models for cometary H~II regions 
can be distinguished by the velocity structure of the ionized gas.
Several studies have been carried out using radio recombination lines
to map out the velocity structure. Garay \etal\ (1986) concluded that the
cometary region in the G34.3+0.2 complex exhibits a champagne flow,
but also shows a large velocity gradient perpendicular to the symmetry
axis which they attributed to rotation of the parent cloud. A lower
resolution study of the same object by Gaume, Fey
\& Claussen (1994) found a similar velocity structure, but they reject
both bow-shock and champagne flow models in preference to a picture
involving interactions with outflows from other sources ahead of the
bow. Similar velocity structures have been seen in two cometary
regions in the Sgr B2 complex by Gaume \& Claussen (1990). Garay
\etal\ (1994) presented radio recombination line maps of
more extended H~II regions and find evidence for a bow-shock in
one and champagne flows in two other more clumpy sources.
Wood \& Churchwell (1991) carried out high
resolution radio recombination line studies of the cometary UC H~II region
G29.96--0.02. Their results were analysed by Van Buren \& Mac Low (1992), who
claim good agreement with their bow-shock model.  

There are substantial problems with the radio recombination line
approach.  These lines are intrinsically very weak, and since they
arise from levels well above the ground state (Wood \& Churchwell 1991
used H76$\alpha$), are prone to many broadening effects and maser
activity. The high spatial resolution interferometric observations
only detect the lines near the head of the bow and resolve out the the
weaker and more diffuse tail emission. The synthesised beam sizes
required to detect the radio recombination lines in the tail are
usually then too large to resolve the head region simultaneously.  By
contrast, infrared recombination lines are intrinsically brighter, and
the detector technology is such that we can map emission across much
larger regions where the surface brightness of the line emission is
lower at high and uniform spatial resolution. Even allowing for the
large extinction in these objects (A$_{V}\sim20-30$ is typical, and
much larger values possible), mapping the HI Br$\gamma$ or Br$\alpha$
emission provides many potential benefits over further radio surveys.

We have therefore embarked on a series of IR observations of UC H~II
regions selected from Wood \& Churchwell (1989) and from Kurtz, Wood
\& Churchwell (1994).  In this paper we demonstrate the value of our
method by presenting observations and analysis of the prototypical
cometary UC H~II region: G29.96-0.02.  In future papers we will
present further observations.

\section{OBSERVATIONS}
We used a broad band K$_n$ image obtained with the near-infrared
camera/spectrometer IRIS on the AAT to determine slit positions 
on the target.  These positions are shown in Fig.\ \ref{fig:slits},
overlaid on a contour map of the K$_n$ image.  
The image has been rotated by 120$^\circ$ east from north.
The near-infrared morphology
of G29.96--0.02 is similar to the radio data presented in, for example,
Afflerbach et al.\ (1994) or Fey et al.\ (1995).  
We will refer to the `head' of the region
as being that portion at the top of Fig.\ \ref{fig:slits}, whilst the
`tail' is the broad fan shape at the bottom.
Also of note in the contour map is the bright star located 2$''$
behind the `head' of the HII region, which we identify as a possible
candidate for the exciting star for the region.  Other stars are also
evident in the field of view, but these do not appear to have associated
ionised gas.  We also obtained `snapshot' images at J and H as well
to determine the nature of this star.

The high resolution spectra were obtained with the common user IR array
spectrometer CGS4 (Mountain et al.\ 1990) on UKIRT on the night of 23
June 1994.  We used the echelle grating within
CGS4, with a one pixel wide slit (each pixel corresponded to 1.1$''$
in the dispersion direction and 2.0$''$ along the slit).  The
effective resolution of this combination was measured to be
21.2~\kms\ on a completely unresolved bright OH night sky line.  To
fully sample the resolution element, the array was stepped six times
across two pixels.  This oversampling is extremely useful in
determining accurate line profiles for our data.  Our basic observing
technique was to observe at each position for ten minutes (composed of
five two minute exposures), and then take a separate two minute sky
frame.  The slit was set at a position angle of 120$^\circ$, as shown by
Fig.\ \ref{fig:slits},  in order
to be approximately along the axis of \gtn\ and we stepped 2$''$
between each position (so that we are undersampled spatially).

The echelle in CGS4 uses a circular variable filter (CVF) as an order
sorter.  This CVF can introduce significant fringing into the final
image.  To compensate for this we obtained `fringe-frames', by
observing the flat field lamp in the same manner as we observed the
objects.  Since our object is extended like the lamp, dividing all
frames by this pattern removed virtually all trace of the fringing.

The wavelength region observed contains only one bright sky line, and a few
much weaker ones.  To remove the night sky emission, we used the median of all
the separate sky frames, and subtracted this from each two minute object
exposure.  The two minute exposures for each position were then averaged.
One of the features of the echelle optics used in CGS4 is that the projection
of the slit onto the array is curved.  We removed this curvature by obtaining
distortion corrections based on the strong night sky emission line at
2.17114~$\mu$m (Oliva \& Origlia 1992:  note, all wavelengths are quoted
in vacuum) from our median sky frame.  

To wavelength calibrate our spectra we used an internal argon lamp within CGS4.
There is a strong line at  2.167561~$\mu$m (Humphreys 1973) which was visible
within our wavelength range.  We assume a purely linear shift in the wavelength
scale  (which we have been assured is reliable for the echelle grating by the
UKIRT staff).  For the pixel with the largest flux (at offset (--2,+2.1)),
we find $v_{\rm LSR}$=95$\pm5$~\kms, where the error reflects uncertainties in
the correction to LSR velocities,  in our measured wavelengths and in the
chosen rest wavelength for hydrogen (we assumed a wavelength of 2.166125~$\mu$m
using the values for energy levels given in Bashkin \& Stoner 1975).
For the same position (that of peak radio flux), Wood
\& Churchwell (1991) quote a velocity of 98~\kms\ (no error is given), and
Afflerbach et al.\  (1994) quote $(92-96)\pm4$~\kms\ for a similar position and
beam size to ours.  If we use the OH night sky line to calibrate the data
we would have a velocity that was some 20~\kms\ too large.  We therefore
find that the value quoted for the wavelength of this line is inaccurate to
this amount.

Observations were also made of BS7377, a F3IV star with $K=2.52$,  which served
both as a flux and atmospheric standard.   In practice, it proved impossible to
remove all the CVF fringes from this object (probably due to the star being
effectively an unresolved source in the slit, whereas the flat field correction
is a diffuse illumination).  We checked all our spectra to ensure
that the line emission was in a region free of significant atmospheric
absorption, rather than actually dividing by this standard.  We used the
relative counts in object and standard to flux calibrate our data.   For the
same pixel we compared the wavelength calibration for, this results in a 
Br$\gamma$ flux
of $5.9\times10^{-16}$~\Wm.  Our maps can be calibrated from this value
using figure 7. By fitting the weak Br$\gamma$ absorption line in this
standard, and using the known radial velocity, we derived another check on the
wavelength calibration. We found this calibration gave a wavelength for the
pixel with the largest flux
of 105$\pm$10~\kms.  This agreed with the calibration using the 
arc line (within the errors, which are mostly due to the CVF fringing).  We are
therefore confident of our wavelength calibration  within the error quoted.
We also used the standard star observations to monitor the seeing:
we estimate that throughout these observations it was $1-2''$.

\section{RESULTS}
We obtained very short `snapshot' images (typically 10 seconds on object and
10 seconds on sky) in J, H and K$_n$ using IRIS to
estimate the magnitude of the putative exciting star.  From these, we find
K$_n$=11.2, H=12.8 and J=15.2 (with typical errors of $\sim0.3-0.5$ magnitudes,
due to (i) the short exposures and (ii) the difficulty in separating the
stellar component from the free-free background of the nebula).
We
can convert these into absolute magnitude given the extinction and the distance
to the source.  By comparing the observed Br$\gamma$ flux in a large beam with
the radio flux at a sufficiently high frequency that the radio continuum is
optically thin we can estimate $\tau_{Br\gamma}$ (cf Lumsden \& Puxley 1995).
There are two published values for large aperture Br$\gamma$ fluxes: Doherty et
al.\ (1994) found a flux of $2.5\times10^{-15}$~\Wm\ in a 5$''$ aperture
centred on the continuum peak; Herter et al.\ (1981) found a flux of
$7.6\times10^{-15}$~\Wm\ in a 12$''$ aperture; and lastly summing over
our sparsely sampled emission line map (Fig.\ \ref{fig:greyf}) we find a flux of
$7.3\times10^{-15}$~\Wm\ over the whole source.  We believe the latter two
values are more representative of the actual total flux in the HII region, and
use the Herter et al.\ value below since it does not suffer from the sparse
sampling problems our CGS4 data does.  There are two high frequency radio
measurements with suitable spatial resolution: Wood \& Churchwell (1989) find
that the 15~GHz integrated flux density is 2.661~Jy; Cesaroni et al.\ (1994) find
that the 24~GHz flux density is 3.534~Jy.  We prefer to use the latter value
since the VLA configuration used gave slightly greater sensitivity to the
extended emission that is clearly missing from the Wood \& Churchwell data
(compare our K$_n$ map with their 15~GHz one for example), and also because the
continuum opacity will be even less of a problem at 24~GHz.  Assuming an
electron temperature of 7500~K (which is consistent with the scatter
in the values found by Afflerbach et al.\ 1994), 
and using the appropriate Br$\gamma$ emissivity
from Hummer \& Storey (1987), this gives a
total expected Br$\gamma$ flux of $5\times10^{-14}$~\Wm.  Hence, we find
$\tau_{Br\gamma}=1.9$.  Using the extinction law of Landini et al.\ (1984),
this corresponds to $A_K=2.0$, $A_H = 3.4$ and $A_J = 5.7$ (and by further
extrapolation $A_V=24$).  These values compare extremely well with those
derived from a comparison of the observed and expected near--IR HI line
strengths (Toby Moore, private communication).
For an assumed distance of 7.4~kpc (Churchwell,
Walmsley \& Cesaroni 1990), we therefore derive absolute magnitudes for the
star(s) of $M_K=-5.1$, $M_H=-4.9$ and $M_J=-4.8$.  Comparison with the colours
expected of main sequence O stars (Koornneef 1983) show reasonable agreement
with the relative JHK values within the errors given above.  From the expected
optical-infrared colours we derive $M_V\sim-6.0$, which is 
typical of an O3V
star.  This value is consistent with that obtained from the radio
and far infrared data (Wood \& Churchwell 1989 give the type as O4--O5,
which have $M_V=-5.7\rightarrow-5.9$).  
However, given the absence of strong 10.5~$\mu$m [SIV]
emission in \gtn\ (eg.\ Herter et al.\ 1981), which we would expect to be
present for a single star of this type, it is likely that the exciting source
is in fact a cluster of stars. The hottest star is likely to be of type O6 at
most with this constraint (see also the arguments presented about the very
similar compact HII region, G45.12+0.13, in Lumsden \& Puxley 1995).

All our CGS4 spectra were fitted by a single Gaussian line profile using the
Starlink {\small DIPSO} package.  We allowed
the intensity, central wavelength and line width to vary.  A Gaussian is a
good match to the instrumental line profile (found by fitting
the night sky emission line).  Except where the signal-to-noise was
very low, we found that a Gaussian was generally a good match
to the observed Br$\gamma$ profile.  There is some small amount of
evidence for weak line splitting near the head and the tail (see also
Section 4.1).  There
was negligible continuum emission in any position.

We present the results of our Gaussian fits to the observations in
image form in Figs. \ref{fig:greyf} to \ref{fig:greyw}, where we have
excluded those points with the lowest signal-to-noise (by trial and
error a signal-to-noise of $\sim$20 in the measured flux
appears to be the minimum value
for which the Gaussian line fitting parameters are entirely reliable). 
The flux map is very similar to the K$_{n}$-band continuum contour
map shown in Fig.\ \ref{fig:slits}.  This is not surprising since in
general the continuum emission at the short end of the $K$ window is
dominated by free-free processes and hence line and continuum should
track each other. 

We can also compare our data to that presented by Wood \& Churchwell (1991) and
Afflerbach et al.\ (1994).  The former has the advantage of good spatial
resolution, but poorer sensitivity, whilst the latter has a larger beam but is
more sensitive.  The Wood \& Churchwell map of the line emission covers an area
of about 7$''\times6''$; the Afflerbach map extends $\sim14''$  along
the tail but at much poorer spatial resolution (they had to use a 4.2$''$
beam to sample the gas in the
tail, and clearly smooth over much of the structure present as they note
themselves).  By contrast our data accurately maps the line emission across
$\sim20''\times17''$  as shown in Fig.\ \ref{fig:greyf}.  Our data are
 similar
to the total extent and structure of the radio continuum emission in
the Fey et al.\ (1995) or  Cesaroni et al.\ (1994) continuum maps.  
 We are therefore in a better position to probe the velocity
structure in the important regions near the tail, and around the edges of the
cometary profile.  As shown below, these provide important constraints on the
models.  Allowing for the different beam sizes involved we find excellent
agreement between our data and both the 
Wood \& Churcwell and Afflerbach et al.\ datawhere we overlap.  The most
obvious discrepancies between our data and the Wood \& Churchwell data (which
has received extensive attention) are: (i) the line centre velocity reaches a
maximum near the line intensity peak, and then turns over, such that the
velocities seen near the tail are very similar to those at the head; (ii) the
line widths are broader in all directions away from the intensity peak
and not just towards the head.
 As we show below, taking account of
these features leads us to rather different conclusions from Van Buren \& Mac
Low (1992).

\section{MODELLING}
We have interpreted our velocity data by constructing an empirical
model where the three dimensional density distribution and velocity
structure are specified and the resultant simulated spectra are then
compared to the observed quantities. This approach gives us
flexibility and means we are not restricted to one particular model.

Our methodology resembles in some respects that used by Mac
Low \etal\ (1991). In particular we follow the geometric layout of
their Fig.\ 11 in which our object is specified in a cartesian system
($x^{\prime}y^{\prime}z^{\prime}$) and is axi-symmetric about 
the $z^{\prime}$ axis. The object frame is at an
angle $i$ relative to the observers frame ($xyz$) rotated about the
$x^{\prime}$ axis (parallel to the $x$ axis) and the observer is at $z=+\infty$.
This geometry is illustrated in Fig.\ \ref{fig:geom}.

We investigated the bow shock model first as this is more amenable
to analytic specification. We attempted to use analytic forms to
match closely the numerical models of Mac Low \etal\ (1991) which have
already been compared to the radio recombination line data for \gtn\
by Van Buren \& Mac Low (1992). The shape of the bow shock was assumed
to be described by the function
\begin{equation}
z^{\prime}=r/tan(r/l)
\label{eqn:tan}
\end{equation}
in cylindrical coordinates where $r$ is the radial coordinate parallel
to the $y^{\prime}$ axis. This shape was derived by Dyson (1975) and was shown
to be a good approximation to the numerical solution calculated by Mac
Low \etal\ with the tangent function slightly underestimating the
divergence in the tail of the bow. The function intersects the
$z^{\prime}$ axis at $z^{\prime}=l$, which was termed the standoff
distance by Mac Low \etal~~The electron density distribution as a
function of $\phi$ - the angle with respect to the $z^{\prime}$ axis from
the origin - was described by a gaussian such that 
\begin{equation}
n_{e} = n_{e}(\phi_{0})exp(-\phi^{2}/2\sigma^{2}_{n\phi})
\end{equation}
where $n_{e}(\phi_{0})$ is the density on the $z^{\prime}$ axis.
A value of $\sigma_{n\phi}=1$ closely approximates the distribution
shown by Dyson (1975) in his Fig.\ 5. To describe the thickness of the
shell we again use a gaussian such that
\begin{equation}
n_{e}(\phi_{0})=n_{e}(max)exp(-(l-l_{max})^{2}/2\sigma^{2}_{nl})
\end{equation}
where $l_{max}$ is the point on the $z^{\prime}$ axis where the
density is at a maximum.
Regions where the density fell to less than 0.1\% of the peak were
not considered in the calculation.

The velocity distribution was specified so as to match the form given
by the numerical models of Mac Low \etal\ illustrated in their Fig 3a.
We assumed that the velocity
is perpendicular to the shell (the velocity is defined relative to the
ambient ISM in this case) since that seems to be the case to
within about 10\degs. The fall-off of speed with $\phi$ was assumed
to be of the form
\begin{equation}
v_{bow}=v_{\star}(1-(\phi/c_{\phi})^{2})
\label{eqn:vbow}
\end{equation}
With $c_{\phi}$=3.0 this closely approximates the numerical solution
of Mac Low \etal~~$v_{\star}$ is the velocity of the star relative to
the ambient cloud in the bow shock model. 
The density and velocity distributions are rotated about the
$z^{\prime}$ axis in order to give the 3D structure.

The model was calculated on a cartesian cube in the observer's frame of
101 points on each side with a step size of 0.4$''$ so that the
total size of the cube is $\pm$20$''$ from the origin. 
This was sufficient to sample the
slit width and cover the entire area where emission was detected. We
proceeded through each $x,y$ point in the sky frame in turn and integrated
along the $z$ axis to give the emission measure (EM) at each velocity
such that
\begin{equation}
EM(x,y,v)=\int n_{e}^{2}(x,y,z,v) dz
\end{equation}
To obtain the density and line-of-sight velocity at any one point we
first transformed the coordinates into the object frame (i.e. 
$x^{\prime}y^{\prime}z^{\prime}$) . The equivalent
cylindrical coordinate, $r$, is given by
(${x^{\prime}}^{2}+{y^{\prime}}^{2})^{\frac{1}{2}}$ and then $l$ can be
found from inverting equation \ref{eqn:tan}. With
$\phi$ given by arctan($r/z^{\prime}$) we obtained the density and
velocity components in the object frame. The velocity components are
transformed back into the observers frame to give the $z$ component of
velocity for that point.

For the intrinsic spectrum of the \brg\ line 
we considered only thermal and turbulent
broadening and assumed a gaussian profile given by
\begin{equation}
\phi_{D}(v)=\frac{1}{(2\pi)^{1/2}\sigma_{v}}exp(-(v-v_{z})^{2}/2\sigma^{2}_{v})
\end{equation}
where
\begin{equation}
\sigma^{2}_{v}=\half\left(\frac{2kT}{m_{H}}+v^{2}_{turb}\right)
\end{equation}
We assumed a constant electron temperature throughout the nebula of
6500~K as deduced from the radio recombination line observations of
Afflerbach \etal\ (1994). The turbulent velocity was left as a free
parameter to match the observed width of the lines and was also
assumed to be constant over the nebula. Fifty frequency points
covering $\pm$70~\kms\ were used.

For each frequency the emergent emission measure was then convolved
with a gaussian seeing profile of FWHM=1.0$''$ and then integrated
over the area of each pixel of each of the observed slit
positions. The resultant spectrum was then convolved in the spectral
direction with a gaussian to represent the spectral resolution of the
CGS4 spectrometer. In order to compare with the data, the synthesised
spectra were subjected to the same gaussian fitting process. The model
parameters were varied until as reasonable a match as possible was
obtained.

\subsection{Bow shock models}

The results for our best bow shock model as described above are shown in Figs.
\ref{fig:bowcol} and  \ref{fig:bowrow} with the model parameters given in Table
\ref{tab:bowres}. Fig.\ \ref{fig:bowcol} shows the comparison of the flux,
velocity centroid and velocity width of the gaussian fits to the model compared
to those for the data along the direction of the slits whilst Fig.\
\ref{fig:bowrow} show the data from the same spectral
row of each slit position i.e.\ the data perpendicular to the slit direction. 
All the fluxes, both observed and model, have been normalized to the peak
observed flux in the central slit position.  
In order to compare the
synthesised velocity centroids with the data we have subtracted a velocity
offset which is another free parameter and is basically the $v_{LSR}$ of the
ambient cloud in the context of the bow-shock model.   

The two curves in Fig.\ \ref{fig:geom} trace the points where the
density has fallen to half its peak value and the total integrated
emission measure from this model is similar to that presented by
MacLow et al.\ (1991) for this source.  As
in the model by Van Buren \& Mac Low the best fitting inclination
angle appears to be $-$135\degs\ (the head of the bow pointing away
from us) and a stellar velocity of 20~\kms\ appears most appropriate.
Our value of the standoff distance $l_{max}$, which is constrained by
the need to fit the opening angle of the structure with equation
\ref{eqn:tan}, is over twice the 1.3$''$
derived by those authors. This is in the sense expected since the
tangent function we use to describe the shape of the bow shock
(equation
\ref{eqn:tan}) underestimates the divergence of the tail. It can be
seen from the fit to the cross-sections in the tail in Fig.\ \ref{fig:bowrow}
that our analytic description still underestimates the width of the
tail somewhat. Increasing $l_{max}$ further does not not improve the
overall fit since this begins to displace the peak in the velocity
centroid too far away from the flux peak. Obviously the tangent
function is not a perfect description of the shape of \gtn.

As an aside, it is worth noting the position of the bright star (or
cluster of stars) discussed in section 2.  The projected standoff
distance is 2.0$''$ which for an inclination close to $-$135\degs\
(45\degs) is close to the deprojected value in the bow shock picture
(see Fig.\ 5 of Mac Low \etal\ 1991). If this were the exciting star,
as argued in Section 3, 
then the inclination angle would have to be closer to $-$150\degs\
(30\degs) in order to agree with the standoff distance derived by Van
Buren \& Mac Low (1992). However, their estimate is strongly dependent on the
stellar parameters used (eg stellar wind terminal velocity and mass
loss rate), and small changes to these mean that 
the observed position of the star is still
compatible with their model.

The thickness of the shell ($\sigma_{ne}$) was set by matching the
width of the peak in the flux distribution in the central slit
position (the top left hand panel of Fig.\
\ref{fig:bowcol}). This is only marginally resolved at our
spatial resolution so we also used the FWHM of the cuts
through the radio continuum emission (see Wood \& Churchwell 1989 Figs 13
\& 14) as a constraint. The chosen inclination angle and the angular density parameter
($\sigma_{n\phi}$) also significantly affect the width of the flux
peak as well as the overall shape of the flux distribution along the
central slit position. There is a certain lack of uniqueness among
these parameters and the inclination is determined to no better than
about $\pm$15\degs. However, these parameters do not greatly influence
the pattern of the predicted velocity parameters which is of 
greater importance here.

It can be seen from the first two columns of Fig.\ \ref{fig:bowcol} that
the bow shock model is a very good fit to the velocity centroid
pattern near the head of the bow. The model is also a reasonable fit
to the line widths in the region from the peak of the line emission
towards the tail.  A turbulent velocity of 8.5~\kms\ was found
necessary to explain the minimum width we see in the data of about
30~\kms\ or 22~\kms\ deconvolved.  However, the model is completely
unable to explain the very steep increase in the line widths ahead of
the emission peak.  Here the lines-of-sight are tangential to the shell,
and because of the required orientation all the velocities are parallel
to each other and hence the lines should be relatively narrow.  In the
bow-shock model the highest intrinsic line widths arise 
near the stagnation point of the flow ahead of the bow, where the
gas is streaming in all directions away from that point,  but for the
orientation of \gtn the projected line width from this  effect is small.

From the IR data alone it might be
supposed that these apparently high line widths are due to the
observed spectrum being dominated by light scattered off dust in a
dense shell of neutral material ahead of the bow. This would perhaps
also explain the excess line flux observed ahead of the bow compared
to the model that can be seen in the top row of
Fig.\ \ref{fig:bowcol} as well as in the K$_n$ band image in Fig.\
{\ref{fig:slits}. However, the radio recombination line data of Wood \&
Churchwell (1991) show exactly the same pattern of steeply increasing
line widths {\sl ahead} of the peak emission from the bow (see their
Fig.\ 3). Since the radio data cannot be affected by scattered light
this is a real feature of the direct emission and an alternative
explanation must be found. Perhaps the most likely is that there is a
partially ionized, highly turbulent layer ahead of the main bow shock
itself. Indeed, in both the bow shock and champagne models there is
predicted to be a shock front ahead of the ionization front which
could increase the turbulent velocity in this region.

Another noteable difference between the model and the data lies in the increase
in the line widths nearer the tail towards the edges of the nebula.  This may
be due to simple fluid instabilities in largely turbulent ionised gas (the
passing star in the bow shock model largely evacuates the H~II region, and
therefore those regions well back in the `tail' will slowly fill in as the
ionised and neutral gas at the molecular cloud interface merges).  However,
the current model as it stands is unable to explain this difference.

 Further major deficiencies of the bow shock model can be seen in the
regions away from the head. As the slit position is displaced further
from the axis in Fig.\ \ref{fig:bowcol} the predicted velocity
centroid becomes almost zero along the whole length of the object
whereas the observed points have a similar pattern to the central slit
positions. Obviously in this model most of the motion is directed radially
outward in the tail and so for slits close to the edge this is
predominately tangential to the line-of-sight. A further illustration
of this is seen in the velocity centroid cross-sections in the tail in Fig.\
\ref{fig:bowrow}. The observed velocity centroids
stay red shifted right across the object until finally dropping towards
the rest velocity outside of the well-defined rim at the edge in the
south. The bow shock model necessarily has to follow the cosine-like
fall off towards the edge. Some of this disagreement may be due to the
fact that our assumed geometry is not a perfect match to that
observed, but we believe that this is a major shortcoming of the bow
shock model.

It is also difficult to reconcile the observed difference between the
molecular and ionised gas velocities.  Our velocity offset parameter
of 82 \kms, which represents the LSR velocity of ambient molecular
cloud, is much less than the value of $\sim98$~\kms\ given by
Cesaroni et al. (1991) for various species.  Another way to see this
is that the peak velocity in the ionized gas, $100\pm5$~\kms\ from
this work and 105 \kms\ from Wood \& Churchwell (1991), is less than
10 \kms\ higher than the molecular gas whereas the orientation and
stellar velocity required in the bow shock model predict about 15
\kms.

As well as comparing the data and models through gaussian fits we have
also inspected the data and models for deviations from this line
shape. In the observed line profiles there is evidence of a slight
excess on the red wing right across the brighter areas of the tail.
Non-gaussian profiles are also seen in the model in this region (where
lines-of sight cut through both the advancing and receding faces in
the tail), however in the opposite sense such that there is an excess
on the blue wing and a deficit on the red.

\subsection{Champagne models}

The presence of a velocity gradient along the edge of the cometary
structure implies that the motion cannot be primarily radial but that
there is a component parallel to the object axis as well. This is the
main feature of the champagne model where ionized gas flows towards the
low density direction.  There are only a few models currently in the literature
that treat the more realistic case of an H~II
region expanding into a power-law or exponential density gradient
(eg.\ Franco et al.\ 1989).
We
have only attempted a very simplified prescription which
incorporates the most basic features of the champagne model. To provide
a component of motion parallel to the axis we incorporated a component
$v_{champagne}$ given by
\begin{equation}
v_{champagne}=\left(\frac{z^{\prime}-l_{max}}{z^{\prime}_{max}-l_{max}}\right)v_{champagne}(max)
\end{equation}
in the direction  $z^{\prime}=-\infty$. 
This component obviously
accelerates linearly from the standoff distance to the tail, in reasonable
agreement with the results of the detailed models for similar
smoothly varying density gradients (Franco et al.\ 1990).
At the head of the cometary structure the motion 
still appears to be 
dominated by the expansion of the ionization front perpendicular
to the shell in many of the champagne models.
This can also be seen in the velocity field of 2D hydrodynamical
calculations (eg., Fig.\ 2 of Bodenheimer et al.\ 1979).
Of course this is a similar velocity pattern to that in the bow shock
model. 
Therefore in our champagne models we also retain a component of the form
of equation \ref{eqn:vbow}, albeit with a lower $v_{\star}$ which
mimics the expansion speed of the ionization front. These two
velocity components are added together to produce a total velocity
field which contains the basic points of a champagne model.

The results for a champagne model which has all the other parameters the
same as our bow shock model, except that $v_{\star}$=10~\kms\ and
$v_{champagne}(max)$=20~\kms, are shown in Figs. \ref{fig:blicol} and
\ref{fig:blirow}. We
have also increased the offset velocity by 10~\kms\ to 92 \kms.  This
now represents the $v_{LSR}$ of the exciting star and the ambient
molecular cloud since they are not in significant 
relative motion in the champagne
scenario. Hence the offset velocity now agrees much better with the
observed molecular cloud velocity. As expected similar motions are
seen at the head as in the bow shock model, but now we see a steep
velocity gradient in the tail that is present in all the slit
positions in Fig \ref{fig:blicol}. Of course, we can still not produce
the decrease 
in velocity centroid ahead of the emission peak at the
outermost slit positions. The champagne velocity field also produces
increased line widths in the tail for the central slit positions which
is more at odds with the data than the bow shock model. By contrast, the
same increase in line width towards the edge of the nebula in the tail 
means that those  points  are a better
fit to the data than the bow shock model.  Just as for the bow shock model, it
is possible that turbulence could have a major impact on the line widths
in this region in the champagne model (for example, gas streaming past 
clumped gas would give rise to turbulence).  In
Fig.\ \ref{fig:blirow} the champagne model has slowed the fall-off of
the velocity centroid towards the edges, but still not enough to match
the data.

It is worth considering how well the champagne flow model we have just
described compares with the hydrodynamical models in the literature.
For example, Yorke et al.\ (1983), present radio
continuum maps for champagne models of rather older H~II regions.
These models produce a somewhat more `filled in' appearance than the
observed limb-brightened structure seen in \gtn\ and most other
cometary H~II regions.  The velocity structure of these models is
presented in Yorke \etal\ (1984) (e.g. see their Fig.\ 1).  The
velocity centroid peaks just behind the intensity peak along the
symmetry axis and retains a steep gradient towards the tail at large
off-axis distances just as in our approximate champagne model.  The
behaviour of our model ahead of the star is of course dominated by the
residual bow-shock component we have added in.  In the original
champagne flow model the velocity field is largely static in this
region.  However, combining a champagne model with elements of a model
in which the gas dynamics near the star is driven largely by a stellar
wind (eg.\ Turner \& Matthews 1984) or by radiation pressure on dust
(eg.\ Yorke \& Kr\"{u}gel 1977) may appear somewhat similar to our
combined bow-shock and champagne flow model.  
More detailed calculations of these mixed models are
required to fully answer these questions however.

\section{CONCLUSIONS}
We have demonstrated the value of our infrared techniques in studying
the kinematical structures of young, heavily extinguished H~II
regions.  The data we present here for G29.96-0.02 clearly shows
features not present in previous radio data that allows us to test the
validity of the two proposed models in much greater detail.  In
particular we have shown that the regions in which the bow shock model
has the poorest fit to the data are not at all well represented in the
radio recombination line data of Wood \& Churchwell (1991).
Therefore, their conclusions, and those of Van Buren \& Mac Low (1992)
and Afflerbach \etal\ (1994), that G29.96-0.02 is completely
consistent with a bow shock interpretation must be questioned.

From our own data, and the analytic approximation to the bow shock model we
have employed, we find that the greatest discrepancies lie near the `head' of
the region, and along the `sides'.  For the `head' of the region, we find the
broadest lines are {\em ahead} of the point at which the line intensity reaches
a maximum, whereas the bow-shock model predicts they should coincide.   This
may be due to turbulent mixing processes in the interface between the ionised
and molecular gas.  We do not therefore consider it a strong argument against
the bow-shock model.  Similar arguments could also be applied to the existence
of this gas in the champagne flow model we have described.

The one aspect of the data that the bow shock model cannot explain away through
an appeal to turbulence is the velocity gradient seen along the outer edge of
the H~II region in the `tail'. Our velocity centroid data show large deviations
from the model. Another way to demonstrate this is to compare the maps of line
flux and velocity centroid.  The bow shock model predicts that the observed
velocities along the outer edge of the comet should be constant since we are
seeing gas which is essentially comoving with the molecular cloud material,
e.g. the lowest contour in the velocity centroid map in Fig.\ 6 of Van Buren \&
Mac Low (1992) follows the outer edge of the emission measure. It can be seen
that the opening angle of the velocity centroid is wider than that of the flux
as was also apparent in the radio data (Fig.\ 5 of Van Buren \& Mac Low 1992).  
In the champagne flow model (e.g. Bodenheimer et al.\ 1979), the velocity
gradient behind the star is easy to explain:  however, other means are required
to produce  the velocity gradient ahead of the star at the edge of the H~II
region.  For both models presented here, it is this aspect of the data
that presents the greatest challenge.

In summary, the motions in the tail of \gtn\ are highly suggestive of
a a champagne flow, although our very simple model is still far from a
good fit to all the data.  
Our results are consistent with the radio continuum maps of Fey et al.\ (1995)
who also argue in favour of a champagne flow model on the basis of
the combined radio and near-infrared morphology of the region.
As advocated by Gaume \etal\ (1994) and
others there is an urgent need to investigate champagne flows which
include the effects of stellar winds since we know the latter exist in
these regions as well as taking more realistic density distributions
into account. This will lead to a better understanding of how young
massive stars affect their natal environments and further star
formation.

\section{ACKNOWLEDGMENTS}
SLL would like to thank the Australian Nuclear Science and Technology
Organisation for providing funding for the trip to UKIRT.  We would also
like to thank Tom Geballe and Tim Carroll for their help in obtaining these
observations, and PATT for the allocation of telescope time on UKIRT.  Lastly,
we would like to thank the anonymous referee for suggestions that helped
in the presentation of our data.

\newpage
\section*{TABLES}
\begin{table}[h]
\begin{center}
\begin{tabular}{lc}
Parameter & Value \\
 & \\
Inclination $i$ & -135\degs \\
Standoff distance $l_{max}$ & 3.0$''$ \\
Angular density parameter $\sigma_{n\phi}$ & 1.2 rad \\
Shell thickness parameter $\sigma_{nl}$ & 0.25$''$ \\
Angular velocity parameter $c_{\phi}$ & 3.0 rad \\
Stellar velocity $v_{\star}$ & 20~\kms \\
Turbulent velocity $v_{turb}$ & 10~\kms \\
Velocity offset wrt $V_{LSR}$ & 82~\kms \\
& \\
\end{tabular}
\caption{Parameters for the bow shock model.}
\label{tab:bowres}
\end{center}
\end{table}

\hspace*{0in}\begin{figure}{\begin{center}
\leavevmode
\hspace*{-0.4in}\epsfxsize=3in
\epsfbox{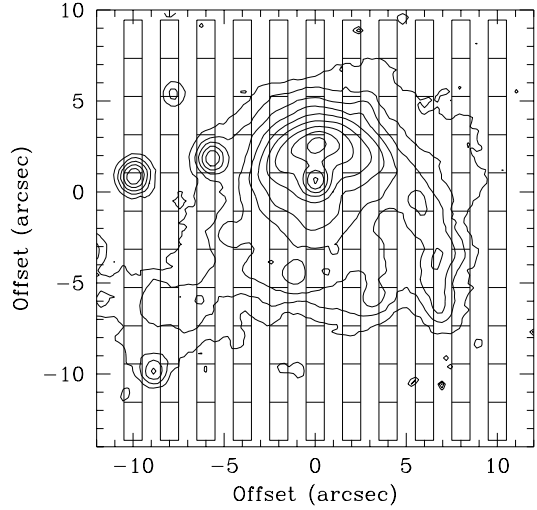}
\end{center}}
\caption{Position of the 11 slit positions used superimposed on\
 a K band contour map of \gtn.  Contours are equally spaced logarithmically
from 3\% to 100\% of peak flux.  }\label{fig:slits}
\end{figure}

\hspace*{0in}\begin{figure}{\begin{center}
\leavevmode
\epsfxsize=3.5in
\epsfbox{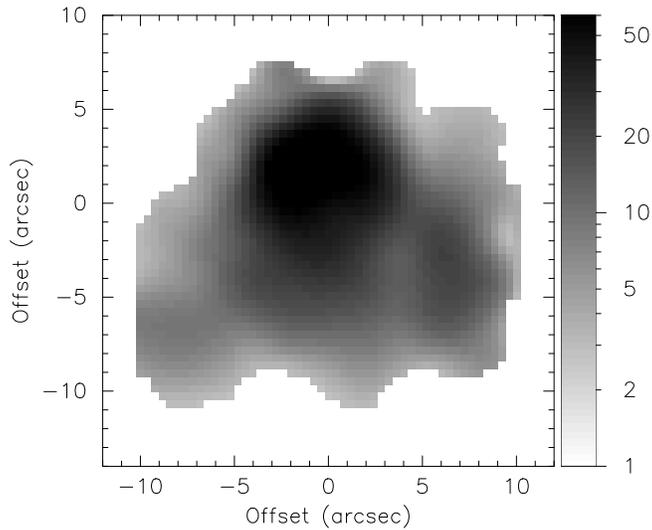}
\end{center}}\caption{Greyscale image of the line flux derived from our
spectra, on the same scale as Fig.\ 1.  The data are plotted logarithmically,
and the scale given is $\log$(F$_{{\rm Br}\gamma}$/$5.5\times10^{-18}$~\Wm).
Sampling is effectively every 2$''$ -- the images are shown on a finer grid for
display purposes only, as are those in Figs.\ 3 and 4.  Only points with a
signal to noise in the integrated flux greater than 25 are shown.  Note that we
have recovered the same structure seen in the K-band image and low resolution
radio maps.  }\label{fig:greyf}
\end{figure}

\hspace*{0in}\begin{figure}{\begin{center}
\leavevmode
\epsfxsize=3.5in
\epsfbox{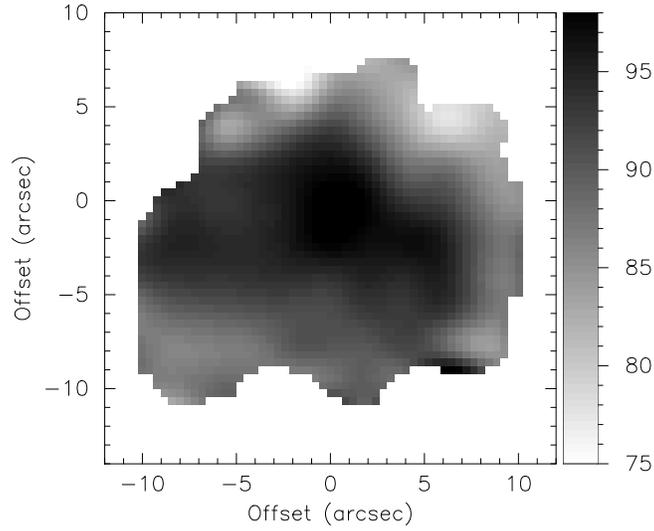}
\end{center}}
\caption{Greyscale image of the velocity centroid derived from
gaussian fits to our spectra. Contour levels are equally spaced from
$v_{\mbox{LSR}}$=75~\kms\ (white) to 98~\kms\ (black). The contours are
perpendicular to the cometary axis in the tail indicating streaming
motion out of the tail.}\label{fig:greyv}
\end{figure}

\hspace*{0in}\begin{figure}{\begin{center}
\leavevmode
\epsfxsize=3.5in
\epsfbox{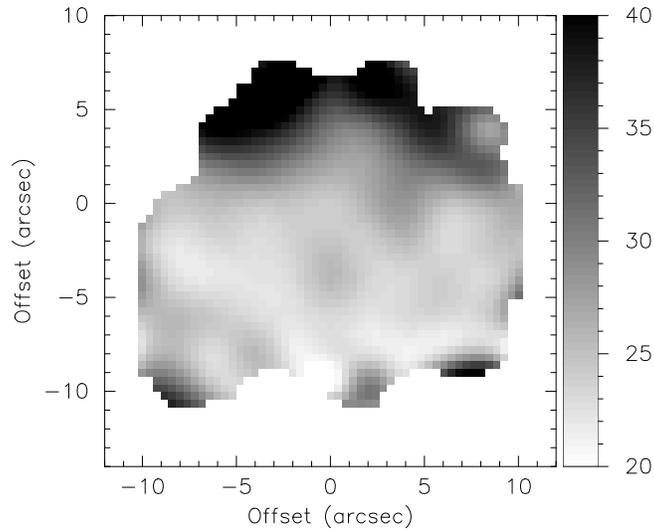}
\end{center}}
\caption{
 Greyscale image of the line width derived from gaussian fits to our spectra.
Contour levels are equally spaced from 20 \kms\ (white) to 40 \kms\ (black).
Note the large line widths ahead of the bow.}\label{fig:greyw}
\end{figure}

\hspace*{0in}\begin{figure}{\begin{center}
\leavevmode
\epsfxsize=3.5in
\epsfbox{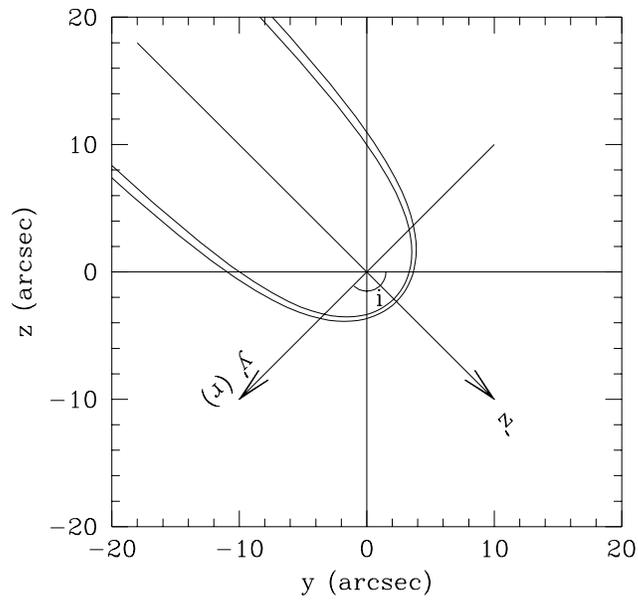}
\end{center}}
\caption{The geometry used in the models showing the definition of
the object with the observer ($xyz$) frame and object
($x^{\prime}y^{\prime}z^{\prime}$, $rz^{\prime}$) frame. The curve traces the
FWHM of the density distribution for the model parameters in Table
1.}\label{fig:geom}
\end{figure}

\hspace*{0in}\begin{figure}{\begin{center}
\leavevmode
\epsfxsize=7.5in
\epsfbox{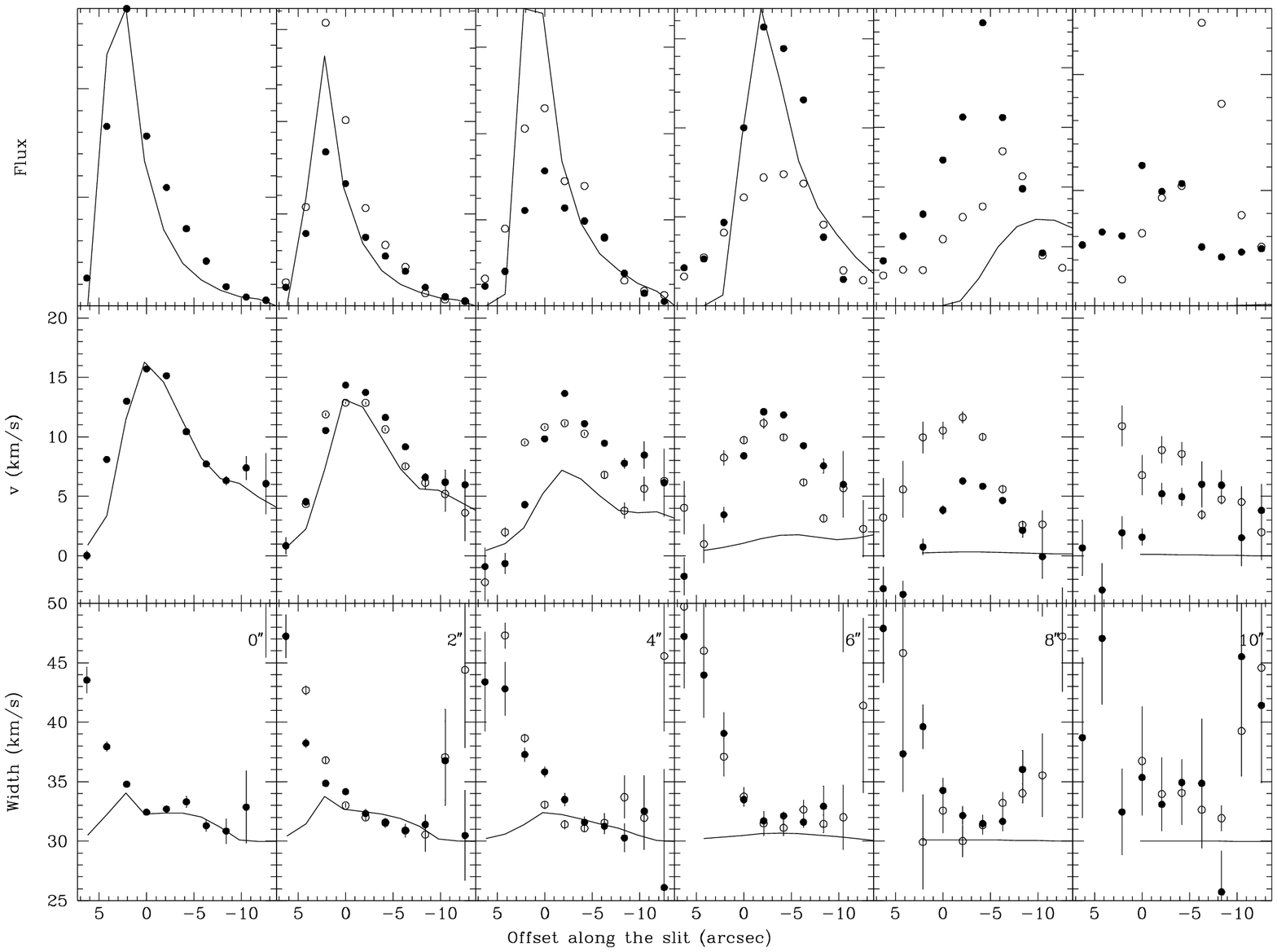}
\end{center}}
\caption{
The flux, velocity centroid and width of the bow shock
model compared with the observational points along the slit. Each set of three
vertical panels (except the first one which shows the on-axis data) shows the
comparison for the two slit positions on opposite sides of the cometary axis
with filled points for the positive offsets and open for the negative ones. The
offset of the slit from the axis is shown in the top right-hand corner of the
bottom panels. Note the lack of velocity gradient in the model compared to the
observations in the slit positions which are a long way from the axis
(right-hand middle panels).}\label{fig:bowcol}
\end{figure}

\hspace*{0in}\begin{figure}{\begin{center}
\leavevmode
\epsfxsize=7.5in
\epsfbox{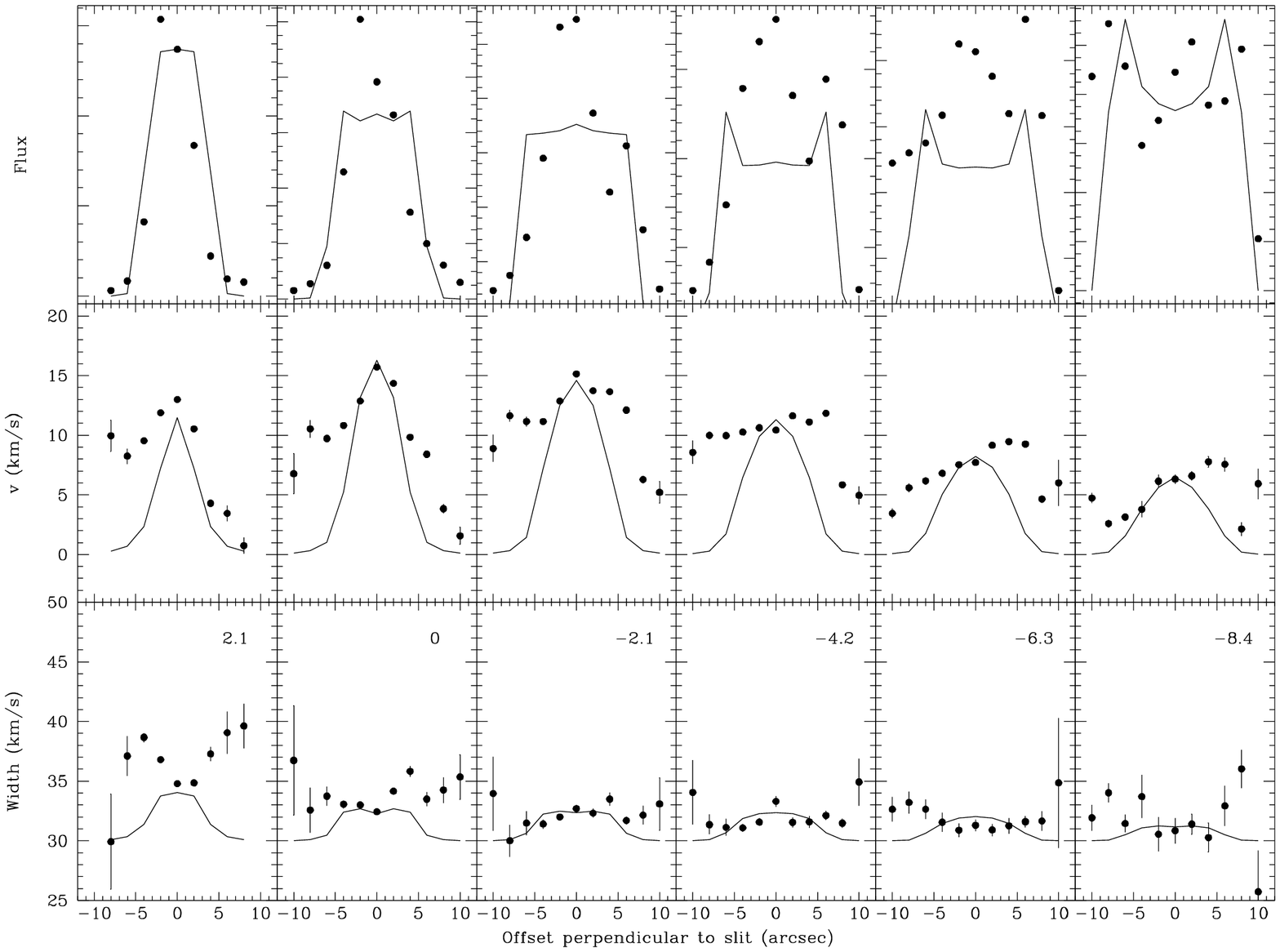}
\end{center}}
\caption{
The flux, velocity centroid and width of the bow shock model
compared with the observational points perpendicular to the slit
direction. Each set of three vertical panels shows the comparison for
one cut perpendicular to the cometary axis. The offsets of the cuts
are shown in the top right-hand corner of the bottom panels and go
from cuts across the head on the left to cuts across the tail on the
right. Note the velocity centroid distribution of the model is too
narrow compared to the observations. }\label{fig:bowrow}
\end{figure}

\hspace*{0in}\begin{figure}{\begin{center}
\leavevmode
\epsfxsize=7.5in
\epsfbox{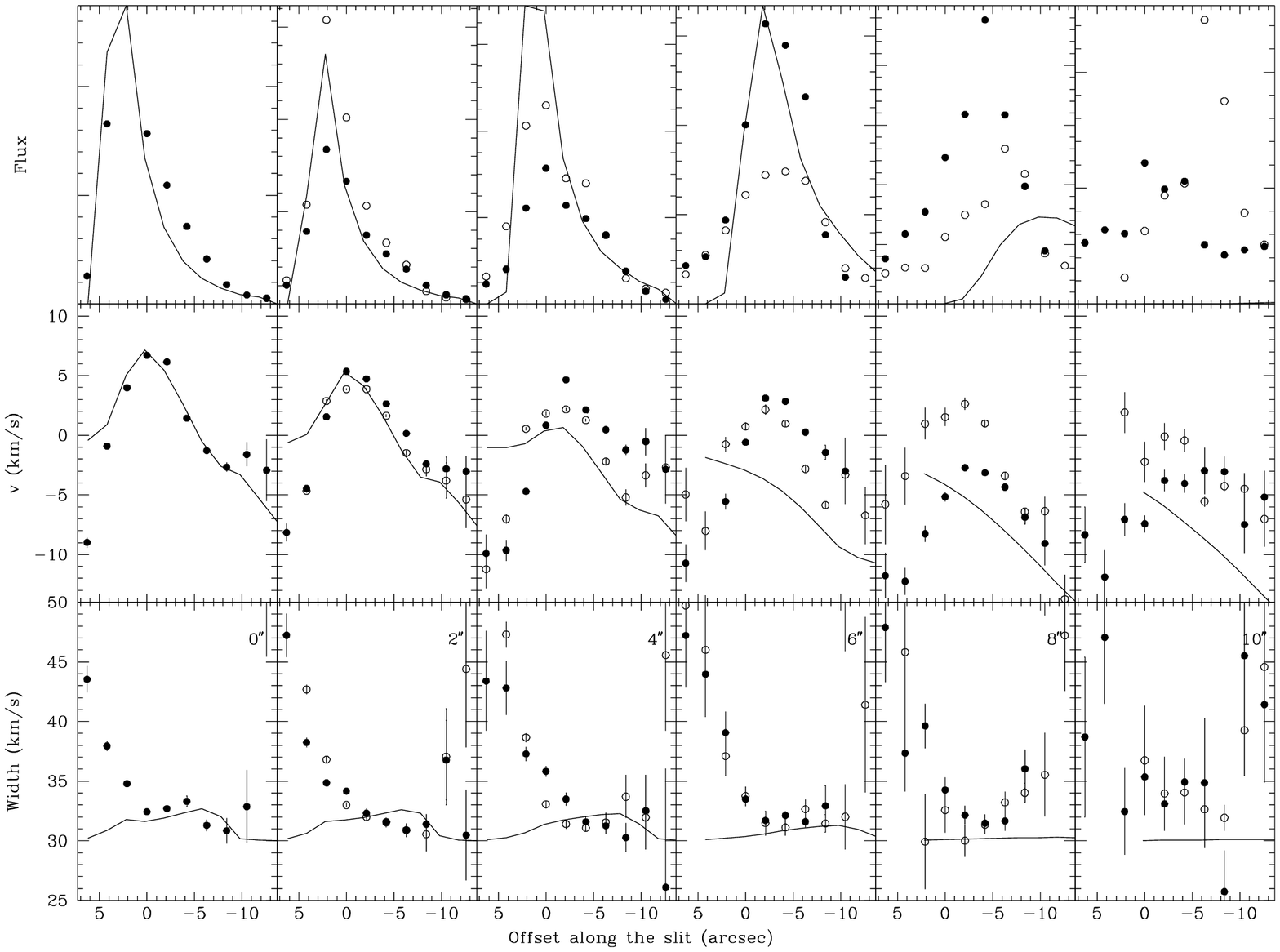}
\end{center}}
\caption{As in Fig.\ 7, but for the champagne model. Note the
model also now has strong velocity gradients in the tail at large slit offset
positions (right-hand middle panels).}\label{fig:blicol}
\end{figure}

\hspace*{0in}\begin{figure}{\begin{center}
\leavevmode
\epsfxsize=7.5in
\epsfbox{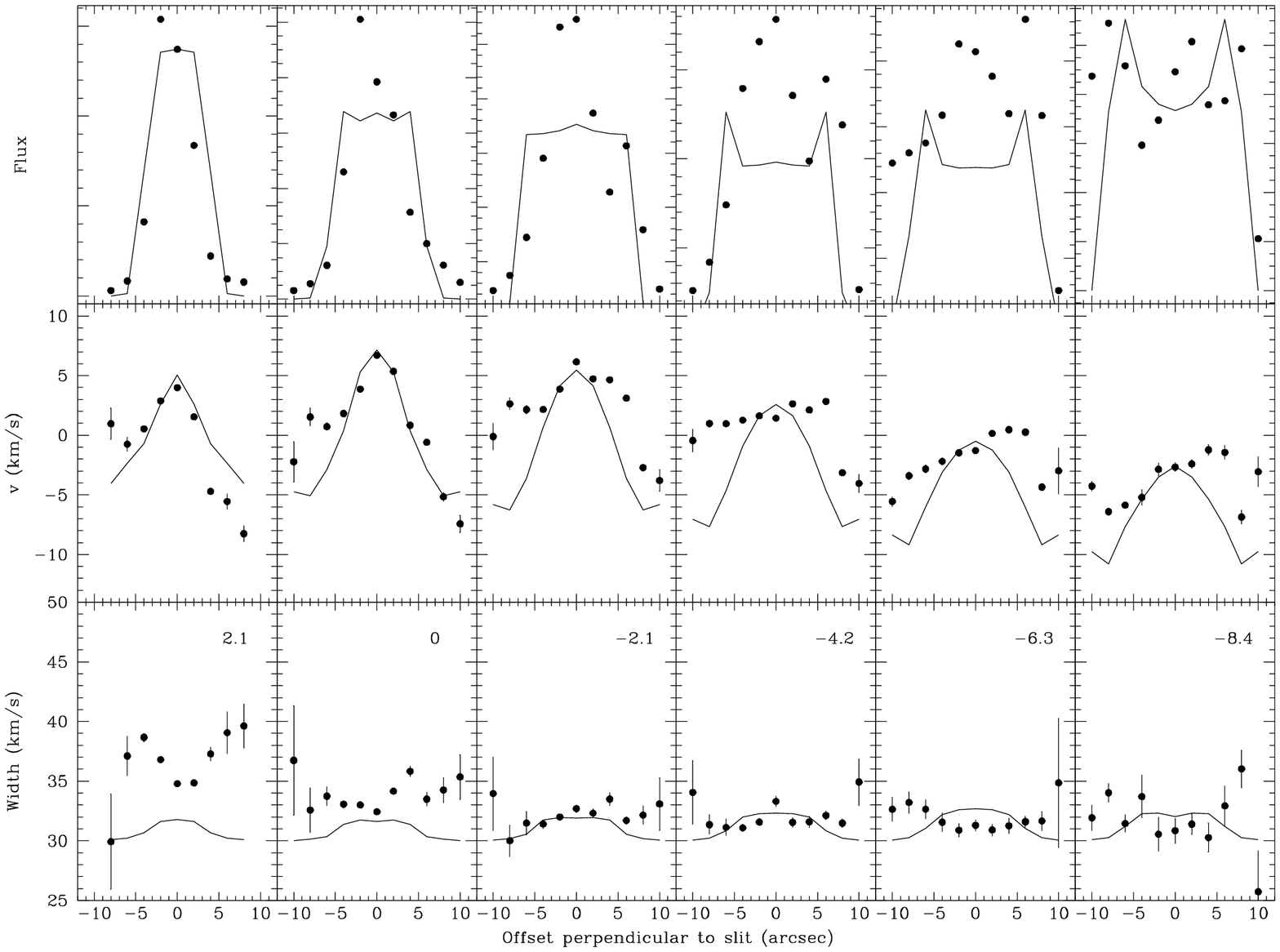}
\end{center}}
\caption{As for Fig.\ 8, but for the champagne model. Note the
velocity centroid distribution is wider and in better agreement with the
observations.}\label{fig:blirow}
\end{figure}

\end{document}